\documentclass[twocolumn,preprintnumbers,amsmath,amssymb]{revtex4}

\usepackage{graphicx}
\usepackage{dcolumn}
\usepackage{bm}

\begin{document}

\preprint{APS/123-QED}

\title{Narrow Line Cooling: Finite Photon Recoil Dynamics}

\author{Thomas H. Loftus, Tetsuya Ido, Andrew D. Ludlow, Martin M. Boyd, and Jun Ye}
 \affiliation{JILA, National Institute of Standards and Technology and University of Colorado, Boulder, CO 80309-0440}

\date{\today}

\begin{abstract}
We present an extensive study of the unique thermal and mechanical
dynamics for narrow-line cooling on the $^1S_0$ - $^3P_1$
$^{88}$Sr transition. For negative detuning, trap dynamics reveal
a transition from the semiclassical regime to the
photon-recoil-dominated quantum regime, yielding an absolute
minima in the equilibrium temperature below the single-photon
recoil limit. For positive detuning, the cloud divides into
discrete momentum packets whose alignment mimics lattice points on
a face-centered-cubic crystal. This novel behavior arises from
velocity selection and ``positive feedback" acceleration due to a
finite number of photon recoils. Cooling is achieved with
blue-detuned light around a velocity where gravity balances the
radiative force.

\end{abstract}

\maketitle

Magneto-optical traps (MOTs) utilizing spin-forbidden transitions
have recently attracted considerable attention as starting points
for all-optical quantum degenerate gases \cite{Takasu03},
single-system tests of Doppler and sub-Doppler cooling theory
\cite{Maru}, and essential components in the next generation of
optical frequency standards \cite{KatoriA, Curtis, PTB}. Of the
currently studied systems, $^1S_0$ - $^3P_1$ Strontium (Sr) MOTs
\cite{KatoriB} are particularly relevant to fundamental atomic
physics since the single-photon recoil frequency shift $\omega_R$
is comparable to the natural linewidth $\Gamma$ and thus
$\omega_R$ directly influences both mechanical and thermodynamic
trap properties. To date, however, many of the rich dynamics for
this unique system remain experimentally unexplored.

In this Letter we report a set of novel $^1S_0$ - $^3P_1$
$^{88}$Sr MOT thermal and mechanical dynamics. For laser
frequencies ($\omega_L$) tuned below the atomic resonance
($\omega_A$), i.e., $2\pi\delta$=$\Delta$=$\omega_L - \omega_A$
$<$ 0, trap dynamics separate into three regimes defined by the
relative size of $|\Delta|$, $\Gamma$, and $\Gamma_E$, where
$\Gamma/2\pi$ = 7.5 kHz and the power-broadened linewidth
$\Gamma_E$ = $\Gamma\sqrt{1+s}$  is determined by the saturation
parameter $s$ = $I/I_S$. Here $I$ ($I_S$ = 3 $\mu$W/cm$^2$) is
the single-beam peak intensity ($^1S_0$ - $^3P_1$ saturation
intensity). Importantly, $\Gamma$ $\sim$ $\omega_R$, where
$\omega_R/2\pi$ = 4.7 kHz. In regime (I), $|\Delta|$ $\gg$
$\Gamma_E$ $\gg$ $\Gamma$ and semiclassical physics dominates.
Photon scattering arises predominantly from single beams over
small, well-defined spatial ranges. Gravity also plays an
important role as the ratio $R$ of the maximum light-induced
acceleration vs gravity $\hbar k\Gamma/2mg$ is only $\sim$ 16,
where $2\pi\hbar$ is Planck's constant, $k$ is the light
wave-vector, $m$ is the $^{88}$Sr mass, and $g$ is the
gravitational acceleration. Trapped atoms relocate to vertical
positions where magnetic-field-induced level shifts compensate
$|\delta|$ and the resultant radiation force balances gravity,
leading to $\delta$-independent equilibrium temperatures. In
regime (II), $|\Delta|$ $<$ $\Gamma_E$, $\Gamma_E$ $\gg$
$\Gamma$, a linear restoring force emerges and thermodynamics
reminiscent of ordinary Doppler cooling including $\delta$- and
$s$-dependent temperature minima occur, although with values
globally smaller than standard Doppler theory predictions. In
regime (III), $s$ approaches unity, the photon-recoil-driven
impulsive force dominates, and the temperature falls below the
photon recoil limit ($T_R$ = 2$\hbar\omega_R$/$k_B$ = 460 nK,
where $k_B$ is Boltzmann's constant) as predicted by a fully
quantum treatment \cite{dalibard89}. The fact that $\Gamma$
$\sim$ $\omega_R$ also enables observations of novel $\delta$ $>$
0 dynamics, where the ultracold sample divides into momentum
packets whose alignment resembles lattice points on 3D
face-centered cubic crystals. This unique behavior, occurring
without atomic or excitation coherence, is first motivated by an
analytic solution to the 1D semiclassical radiative force
equation. Here, we show that for $\delta$ $>$ 0, $\Gamma$ $\sim$
$\omega_R$ allows direct visualization of "positive feedback"
acceleration that efficiently bunches the atoms into discrete,
well-defined momentum packets. The experimentally observed 3D
crystal structure is then shown to arise naturally from the 3D
excitation geometry. In addition, we experimentally demonstrate
that for fixed $s$, $\delta$ determines the lattice point filling
factors, results that are confirmed by numerical simulations of
the final atomic velocity and spatial distributions. More
surprisingly, we find that $R$ directly influences $\delta$ $>$ 0
thermodynamics, enabling cooling around a velocity where
radiation pressure and gravity balance. The physics underlying
this novel behavior is fundamentally the same as regime (I)
$\delta$ $<$ 0 cooling, but manifest in a dramatically different
fashion.

$^1S_0$ - $^3P_1$ traps are formed by first pre-cooling $^{88}$Sr
in a 461 nm $^1S_0$ - $^1P_1$ MOT with an axial magnetic field
gradient dB$_z$/dz (oriented along gravity) of 50 G/cm. The atoms
are then transferred to 689 nm $^1S_0$ - $^3P_1$ MOTs by rapidly
lowering dB$_z$/dz to 3 G/cm and applying red-detuned broadband
frequency-modulated 689 nm light \cite{KatoriB}. Over the next 50
ms, the cloud is compressed by linearly increasing dB$_z$/dz to
10 G/cm. Subsequently, highly stabilized, single-frequency 689 nm
light forms the MOT. The optimal transfer efficiency from $^1S_0$
- $^1P_1$ MOTs to $^1S_0$ - $^3P_1$ MOTs is $\sim$ 30$\%$, giving
final trap populations of $\sim$ 10$^7$. Typical trap lifetimes
and spatial densities are $\sim$ 1 s and $\sim$
5$\times$10$^{11}$ cm$^{-3}$, respectively. Trap dynamics are
monitored either by in-situ or time-of-flight (TOF) fluorescence
imaging.

To gain intuitive insight into trap dynamics, we start with the
semiclassical expression for the force along z,
\begin{eqnarray}
F(v_z, z) = \frac{\hbar k\Gamma}{2}[\frac{s}{1 + s' + 4(\Delta -
kv_z - g_J\mu (dB_z / dz) z)^2/\Gamma^2}\nonumber \\
- \frac{s}{1 + s' + 4(\Delta + kv_z + g_J\mu (dB_z / dz)
z)^2/\Gamma^2}] - mg. \label{eq1}
\end{eqnarray}
where $s'$ ($\geq s$) signifies contributions from other
participating beams and $g_J$ = 1.5 ($\mu$) is the $^3P_1$ state
Lande g-factor (Bohr magneton over $\hbar$). The force along x (or
y) is similar to Eq. (1), but without gravity. Figure 1(a)
presents Eq. (1) for dB$_z$/dz = 10 G/cm, $s$ = $s'$ = 248, and a
range of $\delta$ values. The force is displayed with respect to
position (velocity) in the bottom (upper) axis, for $v_x$ or
$v_y$=0 ($x$ or $y$=0). As $\delta$ decreases, the force makes a
clear transition from the Regime (I) $|\Delta|$ $\gg$ $\Gamma_E$
$\gg$ $\Gamma$ isolated form where excitation occurs over two
separate and well-defined spatial ranges to the Regime (II)
$|\Delta|$ $<$ $\Gamma_E$, $\Gamma_E$ $\gg$ $\Gamma$
dispersion-shaped form wherein excitation occurs over the entire
trap volume and cloud dynamics consist of damped harmonic motion
\cite{JILAsr}. Correspondingly, as $\delta$ decreases, the
initially box-shaped trap potential (Fig. 1(b)) becomes
progressively more ``U"-shaped and the trap shifts vertically
upward. Finally, in regime (III) where $\Gamma_E$ approaches
$\Gamma$ at small $s$, single photon recoils dramatically
influence trap dynamics which in turn requires a full quantum
treatment \cite{dalibard89}.

\begin{figure}
\resizebox{8.5cm}{!}{
\includegraphics{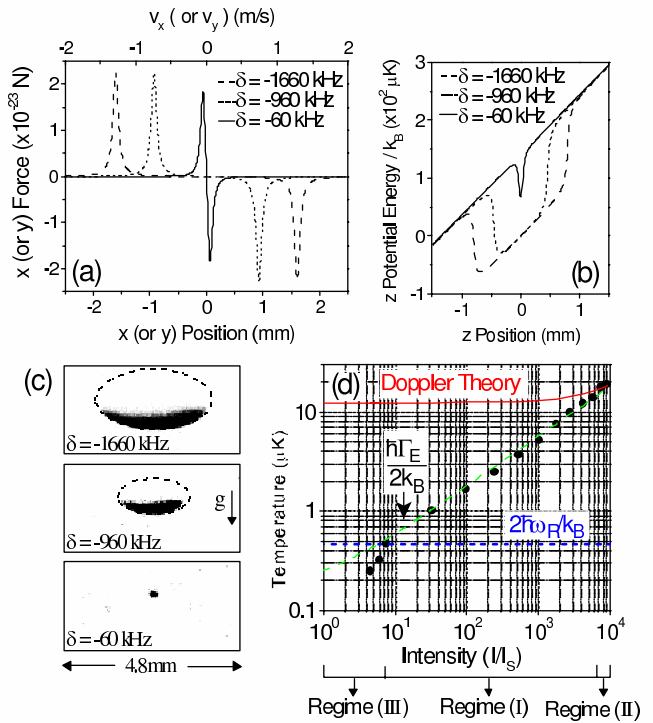}}
\caption{\label{Fig1} (Color online) (a) Calculated radiative
force versus position (bottom axis, $v_x$=$v_y$=0) and velocity
(upper axis, $x$=$y$=0). (b) Trap potential energy in the z
direction. (c) In-situ $^1S_0$ - $^3P_1$ trap images. Dashed
lines are calculated maximum force contours. For each, s = 248 and
dB$_z$/dz = 10 G/cm. (d) Temperature vs intensity for $\delta$ =
-520 kHz and dB$_z$/dz = 10 G/cm. Solid curve: standard Doppler
theory; Long dashed line: Doppler limit ($\hbar\Gamma_E$/$2k_B$);
Short dashed line: single photon recoil limit
(2$\hbar\omega_R$/$k_B$); Filled dots: experimental data.}
\end{figure}

Changes in the force are dramatically revealed in trap mechanical
dynamics (see Fig. 1(c)). In the dispersion-shaped cooling regime
the cloud aspect ratio is $\sim$ 2:1, as expected for a typical
MOT. Conversely, in the isolated regime the atoms move freely
between ``hard wall" boundaries. The cloud horizontal width,
therefore, is largely determined by the separation between
horizontal force maxima, an effect clearly revealed by the
overlaid maximum force contours calculated from Eq. (1).
Moreover, since the radiative force is comparable to gravity
(recall R $\sim$ 16) and the thermal energy is small compared to
the gravitational potential energy, atoms sag to the bottom of
the trap and the lower cloud boundary $z_0$ is well defined by
the point where the Zeeman shift balances $\delta$.

Studying the MOT temperature versus $\delta$ and $s$ provides
rich information about trap dynamics. For large $|\delta|$ and
$s$, corresponding to regime (I), Eq. (1) reflects a balance
between gravity and the radiative force from the
upward-propagating beam at $z_0$ \cite{Com1}. Thus trap
thermodynamics are determined by a Taylor expansion of Eq. (1)
around $v_z$=0 for $z$=$z_0$. With the atomic position ($z_0$)
self-adjusting to follow $\delta$, the damping and momentum
diffusion coefficients are $\delta$-independent, giving a
predicted $\delta$-independent temperature of
\begin{eqnarray}
T(s) = \hbar \Gamma_E/(2k_B)[0.5R(R-s'/s -1/s)^{-1/2}].
\label{eq2}
\end{eqnarray}
We have experimentally confirmed this prediction for a wide range
of $\delta$ \cite{Com4}. The quantity inside the square brackets
is nearly 2, independent of $s$ for the relevant experimental
range. Fig. 1(d) displays the temperature vs. intensity at a fixed
large detuning $\delta$ = -520 kHz, showing good agreement (aside
from a global scaling factor of 2) with the intensity-dependence
given by Eq. (2). This result arises from the semiclassical nature
regime (I) cooling for which $\Gamma_E$ is the natural energy
scale \cite{Lett}.

For regime (II), $|\Delta|$ $<$ $\Gamma_E$, $\Gamma_E$ $\gg$
$\Gamma$, Eq. (1) produces a linear restoring force resembling
ordinary Doppler cooling. Here, we observe \cite{Com4} $\delta$-
and $s$-dependent temperature minima with the minimum and its
$|\delta|$-location both decreasing with $s$. Such behavior is
predicted by Doppler theory, with the ``Doppler Limit" achieved
at $|\Delta|$ = $\Gamma_E$/2. However, in order to match the
data, the theory curves need to be multiplied by a $s$-dependent
global scaling factor ($<$ 1) whose value decreases with $s$.
Moreover, minimum temperatures lie well below the standard
Doppler limit of $\hbar\Gamma_E$/2k$_B$. Notably, this
temperature scaling factor is not explained by semiclassical
Monte-Carlo treatments of the cooling process. In regime (III),
$\Gamma$ $\sim$ $\omega_R$ $\sim$ $k_BT/\hbar$ and the radiative
force acquires a single-photon-recoil dominated impulsive form.
Thus equilibrium thermodynamics can only be adequately described
by quantum theory \cite{dalibard89}. As shown in Fig. 1(d), the
predicted cooling limit of half the recoil temperature $T_R$/2 =
$\hbar\omega_R$/$k_B$ is experimentally reached as $s$ approaches
unity.

\begin{figure}
\resizebox{9cm}{!}{
\includegraphics{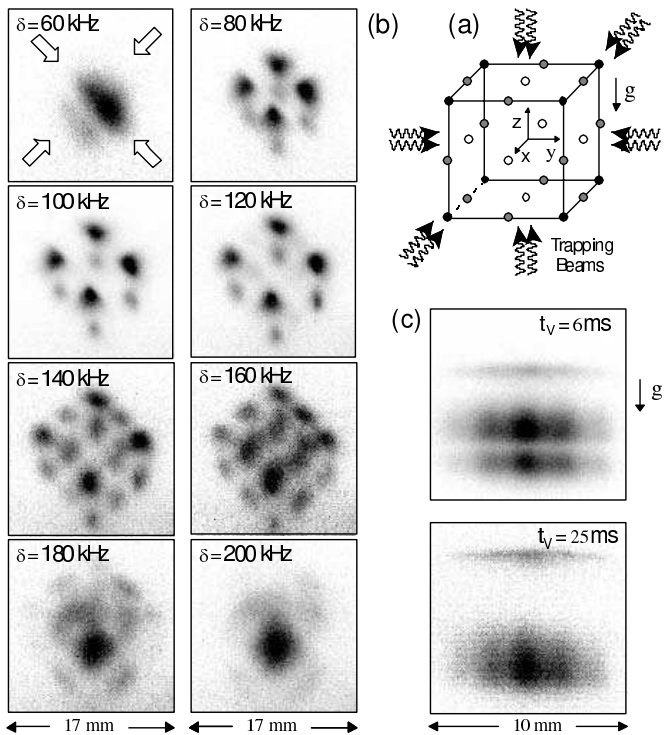}}
\caption{\label{Fig2} (a) Underlying momentum space structure for
$\delta$ $>$ 0. (b) Top view TOF images for t$_H$ = t$_V$ = 25 ms,
t$_F$ = 20 ms. Arrows in the $\delta$ = 60 kHz frame give
horizontal trapping beam directions. (c) Side view in-situ images
for $\delta$ = 140 kHz and t$_H$ = 25 ms. For each, s = 30 and
dB$_z$/dz = 0.}
\end{figure}

Tuning to $\delta$ $>$ 0 presents another intriguing set of
cooling and motional dynamics. Here, the cloud divides into
discrete momentum packets whose alignment mimics lattice points on
a three-dimensional face-centered-cubic crystal \cite{Com2}.
Figure 2(a) depicts the underlying momentum-space structure
which, as shown below, occurs due to highly directional (i.e.,
minimal heating) ``positive feedback" acceleration and velocity
bunching. For the 3D excitation geometry, symmetry dictates that
cube corners correspond to three-beam processes while mid-points
between corners and cube face centers arise from two- and one-beam
processes, respectively. Figure 2(b) shows a $\delta$-specific
sequence of top view (slightly off vertical) TOF images for a
fixed intensity and atom-light interaction time t$_H$ = 25 ms
(t$_V$ = 25 ms) in the horizontal x-y plane (along z-axis)
trapping beams, followed by a free-flight time t$_F$ = 20 ms. All
images are taken with dB$_z$/dz = 0 although we find
qualitatively similar behavior for dB$_z$/dz $\neq$ 0.

At small values of $\delta$ ($\leq$ 60 kHz), the atom cloud
expands nearly uniformly. As $\delta$ increases, 3-beam ''lattice
points" appear first, corresponding to the eight cube corners in
Fig. 2(a). This occurs as the cloud is divided into two oppositely
moving packets along each of the three axes. When $\delta$ reaches
a value around 140 kHz, the atom-light interaction becomes
sufficiently weak for velocities near zero that some atoms remain
stationary along a given axis. These atoms, however, still
interact with the beams along the two other axes causing the
2-beam lattice points to appear. This process forms a total of 20
divided atom packets with 8, 4, and 8 packets present in the top,
middle, and bottom layers of the cube, respectively. As $\delta$
increases further, some atoms are left stationary along two axes,
enabling formation of the 1-beam lattice points, shown as 6 open
circles on the Fig. 2(a) cube face centers. For $\delta$ $>$ 180
kHz, the atom-light interaction weakens further and the original
atom cloud reappears. We emphasize that the temperature
associated with each packet in its moving frame is actually lower
than the t$_V$ = t$_H$ = 0 atomic cloud. This result arises from
the velocity bunching and cooling mechanisms explained below.

Only two vertical layers are observed in Fig. 2(b) while Fig.
2(a) predicts the creation of three. This apparent contradiction
is resolved in Fig. 2(c), where the cloud is viewed in the x-y
plane at 45$^o$ to the x,y axes. In order to explore vertical
dynamics while maintaining evolution in the horizontal plane,
t$_H$ is fixed at 25 ms while t$_V$ is varied between 6 ms and 25
ms. As before, t$_F$ = 20 ms. As shown by the images, the lowest
two layers in Fig. 2(a) are only spatially distinct for short
t$_V$, merging together for t$_V$ = 25 ms. This occurs as gravity
accelerates the middle layer, which is initially stationary along
the z-axis, into resonance with the downward propagating laser
beam. Subsequently the two downward moving layers merge. Hence,
more (less) intense packets in Fig. 2(b) are due to the lowest
two (uppermost) cube layers.

\begin{figure}
\resizebox{8.5cm}{!}{
\includegraphics{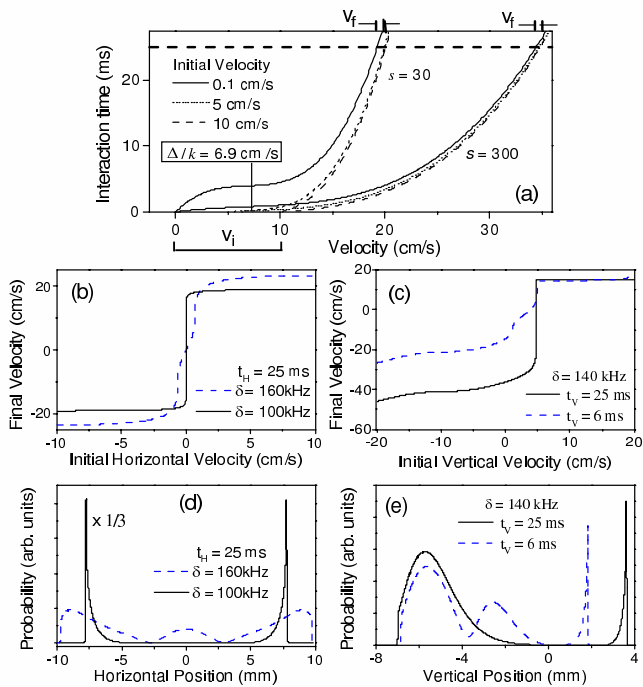}}
\caption{\label{Fig3}(a) Single beam horizontal acceleration and
velocity bunching versus time. Numerically calculated two-beam
final versus initial velocity in the (b) horizontal and (c)
vertical direction. Corresponding spatial distribution in the (d)
horizontal and (e) vertical direction.}
\end{figure}

Quantitative insight into $\delta$ $>$ 0 dynamics can be obtained
from Eq. (1). Recall that for $\delta$ $>$ 0, resonant absorption
occurs between trapping beams and atoms for which
$\vec{\mbox{k}}\cdot\vec{\mbox{v}}$ $>$ 0. The absorption process
thus preferentially accelerates rather than decelerates the
atoms, leading to ``positive feedback" in velocity space that
terminates at a well defined velocity set by $s$ and $\delta$.
For the unique situation where $\Gamma$ $\sim$ $\omega_R$, system
dynamics rapidly evolve toward single-beam interactions. 1D
dynamics can thus be understood by solving Eq. (1) analytically
under a single-beam approximation. The full 3D evolution then
follows naturally from the 3D excitation geometry. Fig. 3(a)
illustrates the evolution of the 1-D atomic velocity versus
interaction time for various $s$ at $\delta$ = 100 kHz. Almost
independent of the initial velocity ($v_i$), the mean value and
spread of the final atomic velocity ($v_f$) are set by $s$ and
$\delta$, which govern how the acceleration process terminates,
leading to efficient velocity bunching. Thus a $\delta$- and
$s$-dependent number of velocity bunched groups are formed.
Considering horizontal motion first, Fig 3(b) shows, for $s$ = 30
and t$_H$ = 25 ms, $v_f$ versus $v_i$ around $v_i$ = 0 for
$\delta$ = 100 kHz and $\delta$ = 160 kHz. In the former case,
atoms at every $v_i$ are bunched into two groups with $v_f$
$\sim$ $\pm$ 20 cm/s. In the latter, three groups appear at $v_f$
$\sim$ 0 cm/s, 23 cm/s, and -23 cm/s. Similar dynamics occur in
the vertical direction where gravity now plays an important role.
Fig. 3(c) shows $v_f$ versus $v_i$ for $s$ = 30, $\delta$ = 140
kHz and t$_V$ times relevant to Fig. 2(c). Notably, for the
upward-moving velocity group, even though $\delta$ $>$ 0, atoms
experience cooling around a velocity $v_0$ where gravity balances
the radiative force, producing the sharp velocity and thus
spatial distribution shown in both the experiment (Fig. 2(c)) and
theory (Fig. 3(c)). Theoretically we find the resultant
equilibrium temperature is given by Eq. (2), as for the
red-detuned case.

For comparison with Fig. 2, Fig. 3(d) and 3(e) give spatial
distributions corresponding to the final velocity distributions
shown in Fig. 3(b) and 3(c), given the measured initial cloud
temperature. Fig. 3(d) corresponds to x or y cube axes in Fig.
2(b). Importantly, the model correctly reproduces cloud shape
asymmetries (see, for example, the sharp edge on the top of the
uppermost layer in Fig. 2(c)), the $\delta$-dependent number of
packets and the relative packet populations, and the
t$_V$-dependent number of vertical layers. Predicted final
velocities and packet spacings, however, are $\sim$ 2$\times$
larger than observed. Measuring the position of the upward moving
layer in Fig. 2(c) versus t$_V$ resolves this discrepancy.
Measured values for $v_f$ are slightly reduced due to small stray
magnetic field gradients that shift $v_0$ as the atoms move
upward \cite{Com3}, giving an apparent downward acceleration.
When these effects are taken into account, predicted and measured
positions agree. Finally, we note that $\delta$ $>$ 0
momentum-space crystal formation is a universal feature of
Doppler limited cooling \cite{Com4}. For broad line cases where
$\Gamma$/$\omega_R$ $\gg$ 1, however, creating structures similar
to Fig. 2(b) requires laser beam diameters on the order of tens
of centimeters and imaging light with hundreds of megahertz
bandwidth, making experimental observations impractical.

In summary, we have performed detailed studies of the transition
from semi-classical to full quantum cooling, revealing signatures
of each regime without ambiguity. Our results show, for the first
time, that the cooling limit of $T_R$/2 can be reached. More
surprisingly, when $\delta$ $>$ 0, the cold atom sample divides
into well defined momentum packets and cooling is achieved around
a velocity where gravity balances the radiative force.

We thank A. Gallagher and J. Hall for useful discussions. This
work is funded by ONR, NSF, NASA, and NIST.


\begin{thebibliography}{10}
\bibitem{Com1}
    For the relevant $\delta$ range, the atoms do not
    absorb photons from the downward-propagating beam (see Fig
    1(b)). Due to polarization considerations, horizontal beam
    absorption rates are smaller than the vertical rate by
    4$\times$.
\bibitem{Com2}
    This analogy is not rigorous since face-centered-cubic
    crystals lack the lattice points along corner connecting lines
    that are observed in the experiment.
\bibitem{Com3}
    Gravity is not essential to this argument; a similar
    effect occurs in the horizontal plane.
\end{thebibliography}
\end{document}